\documentclass[a4paper,fleqn]{cas-dc}

\usepackage{makecell}
\usepackage{graphicx}
\usepackage[numbers]{natbib}

\usepackage{listings}
\usepackage{xcolor}

\usepackage{adjustbox}
\usepackage{float}
\usepackage[ruled]{algorithm2e}
\usepackage{algorithmic}
\usepackage{multirow}
\usepackage{amssymb}
\usepackage{bbding}
\def\tsc#1{\csdef{#1}{\textsc{\lowercase{#1}}\xspace}}
\tsc{WGM}
\tsc{QE}
\tsc{EP}
\tsc{PMS}
\tsc{BEC}
\tsc{DE}
\begin{document}
\let\printorcid\relax
\let\WriteBookmarks\relax
\def\floatpagepagefraction{1}
\def\textpagefraction{.001}

%\shorttitle{Leveraging social media news}

%\shortauthors{CV Radhakrishnan et~al.}

\title [mode = title]{The effect of intelligent monitoring of physical exercise on executive function in children with ADHD}                      

\author[1]{Liwen Lin}
\ead{lwlin@bjtu.edu.cn}
\credit{Conceptualization, Methodology, Writing–original draft}

\author[2]{Nan Li}
\cormark[1]
\ead{nanl@cupl.edu.cn}
\credit{Writing-review \& editing, Conceptualization}

\author[3]{Shuchen Zhao}
\ead{shuchen.zhao.1996@gmail.com}
\credit{Writing-review \& editing, Methodology}

\cortext[cor1]{Corresponding author.}

\affiliation[1]{organization={Department of Physical Education, Beijing Jiaotong University},
   city={Beijing},
   postcode={100044}, 
   country={China}
}

\affiliation[2]{organization={Department of Physical Education, China University of Political Science and Law},
   city={Beijing},
   postcode={102249}, 
   country={China}
}

\affiliation[3]{organization={Duke University},
   city={Durham},
   postcode={27708}, 
   country={Unite State}
}

\begin{abstract}
Children with ADHD often struggle with executive function (EF) and motor skills, impacting their academics and social life. While medications are commonly used, they have side effects, leading to interest in non-drug treatments. Physical activity (PA) has shown promise in improving cognitive and motor skills in children with ADHD. This study examined the short- and long-term effects of three PA interventions: a specific skill training group (EG1), a low-demand exercise group (EG2), and a control group (CG) over 12 weeks. EG1 showed significant improvements in motor tasks and working memory (15\% improvement, p<0.05), while EG2 and CG showed smaller changes. Long-term PA improved working memory, but short-term PA had limited effects on balance and manual dexterity. These findings suggest that skill training has an immediate impact on motor performance, while more complex motor skills require longer interventions. Smart devices tracked progress, confirming sustained engagement and improvement in EG1. This research highlights PA as a promising non-pharmacological treatment for ADHD, warranting further exploration of its effects on other cognitive domains.

\end{abstract}

% Keywords
% Each keyword is separated by \sep

\maketitle

\section{Introduction}

In modern society, children's cognitive and behavioral problems have received increasing attention, especially the executive function and motor performance problems of children with attention deficit hyperactivity disorder (ADHD), which have become a hot topic in many research fields \cite{bunger2021multimethod,liang2022impacts,liang2022physical, dhelim2023detecting}. ADHD is a common neurodevelopmental disorder, mainly characterized by inattention, impulsive behavior, and hyperactivity. According to global statistics, the incidence of ADHD in children is about 5-7\%, and a considerable number of children show severe cognitive and behavioral dysfunction. These problems affect their academic performance and have long-term negative effects on daily life, social skills, and mental health\cite{van2024relationship,song2023meta,ning2023hyper}.

\begin{figure}
    \centering
    \includegraphics[width=1\linewidth]{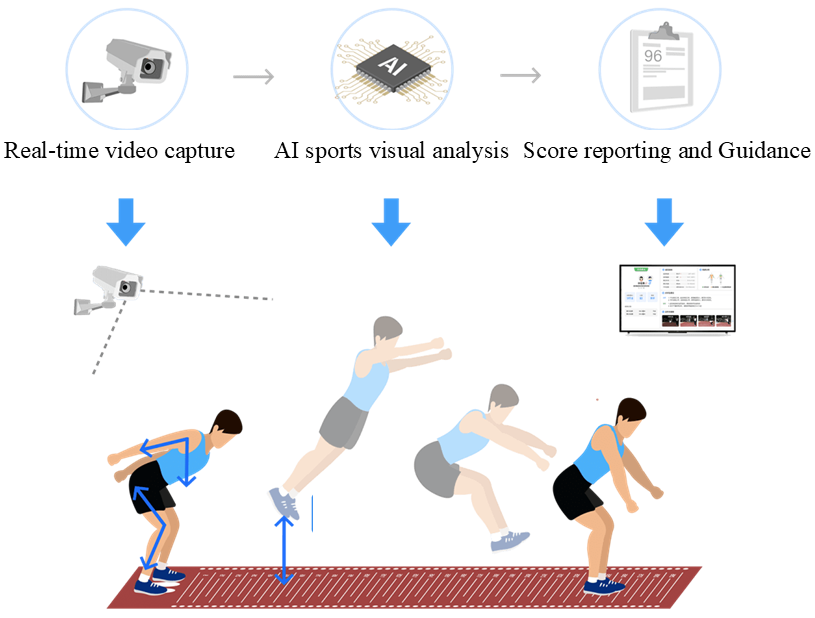}
    \caption{Smart monitoring example diagram.}
    \label{fig7}
\end{figure}

Executive function refers to the ability of an individual to perform a series of cognitive processes such as goal setting, planning, task switching, working memory, and impulse inhibition when facing complex or novel tasks. This ability plays a vital role in children's learning, behavior management, and social interaction. However, children with ADHD often show significant defects in executive function \cite{welsch2021effect,stuhr2020role,wang2025unified,Guo2024}, especially in working memory, task switching, and self-control. Studies have shown that these cognitive deficits not only limit their academic performance but also affect their self-management and social interactions in daily life. In the framework of cognitive psychology ~\cite{andrews2021examining,doebel2020rethinking,spiegel2021relations}, executive function is considered to be a high-level cognitive process. Children with ADHD, often show great difficulties in these cognitive processes, especially in context switching, attention maintenance, and information processing speed. Traditional treatment methods usually use drug interventions, such as methamphetamine drugs (such as Ritalin, Adderall, etc.) ~\cite{suchy2020executive,zhang2024deformation}, which can improve attention and self-control in the short term by regulating dopamine and norepinephrine levels in the brain~\cite{kodipalli2023computational}. However, long-term drug use not only has problems of drug resistance and dependence but may also cause some side effects, such as loss of appetite, insomnia, anxiety, etc. Therefore, more and more parents and psychiatrists are beginning to explore safer, long-term, and effective non-drug treatment methods.

In the non-drug treatment of children with ADHD, behavioral therapy, cognitive training, dietary adjustment, and other methods have been widely used. In recent years, physical exercise (PA) as a positive intervention method has gradually entered the field of vision of researchers \cite{vysniauske2020effects,lund2020adverse}. Physical exercise is not only a simple physical activity but also has a profound impact on the cognitive function of the brain \cite{li2025oral,NING2024102033}. A large number of studies have shown that regular physical exercise can promote neuroplasticity in the brain and enhance the connection between neurons, thereby improving the individual's cognitive ability. Smart monitoring devices have made it possible to track physical activity intensity, duration, and its real-time effects on cognitive function, offering valuable data for fine-tuning interventions based on each child’s progress.

Physical exercise has been widely recognized for its positive effects on executive function in healthy individuals, particularly in areas such as attention, working memory, and impulse control. Research has consistently shown that regular physical activity, especially aerobic exercises, can enhance cognitive performance by increasing blood circulation in key brain regions, including the hippocampus and prefrontal cortex. These areas are critically involved in high-level cognitive processes such as decision-making, problem-solving, and memory~\cite{peng2024automatic,liu2025real,luo2025intelligent,huang2024risk,richardson2024reinforcement,pang2024electronic}. Aerobic exercise has been found to stimulate neural activity in these regions, supporting the neuroplasticity needed for better cognitive function.

Moreover, physical activity plays a vital role in the regulation of neurotransmitters such as dopamine and norepinephrine, which are essential for mood regulation, attention, and focus. By increasing the synthesis and release of these chemicals, exercise not only helps in maintaining emotional balance but also supports the brain's ability to process information efficiently, improve memory retention, and manage impulses. This biological mechanism is fundamental in reducing symptoms of cognitive overload and enhancing the capacity for cognitive flexibility. Furthermore, recent studies have shown that even moderate exercise routines can lead to long-term benefits in terms of brain health, reinforcing the idea that consistent physical activity contributes significantly to both mental and emotional well-being.

Although most studies have focused on healthy individuals, there is some preliminary evidence \cite{zhao2020effectiveness,ludyga2020systematic} that physical exercise can also have positive cognitive effects on children with ADHD. For example, Smith et al. found that after a period of physical exercise, children with ADHD showed significant improvements in attention, task switching, and impulse control. In addition, the study also found that physical exercise can reduce anxiety and emotional instability in children with ADHD and improve their mental health. However, existing studies still have some limitations in methodology. Most studies have a small sample size and lack large-scale randomized controlled trials to verify the generalizability of their results. Therefore, further research is necessary to systematically evaluate the long-term effects of physical exercise interventions on executive function and motor performance in children with ADHD in a larger sample and different types of physical exercise interventions. To further explore the potential effects of physical exercise on executive function and motor performance in children with ADHD, this paper conducted a 12-week intervention study to systematically evaluate the effects of different types of physical exercise programs on cognitive and motor abilities in children with ADHD. This study designed two different types of physical activity intervention programs, an aerobic exercise group and a coordination training group. The study hypothesized that different types of physical activities would have different effects on the executive function and motor performance of children with ADHD. The ADHD children participating in this study were randomly assigned to two groups, each of which underwent a 12-week physical exercise training. During the study, the researchers conducted multiple assessments of the participants' executive function and motor performance, using commonly used cognitive test tools (such as Stroop tasks, task switching tasks, etc.) and motor ability assessment tools (such as coordination tests, strength tests, etc.). In addition, the study also used neuroimaging techniques such as electroencephalogram (EEG)~\cite{sui2024application,xu2022dpmpc,chen2024enhancing,yan2024application,jin2025rankflow,tang2025real,li2025decoupled,shojaee2025federated,ji2024nt,luo2023model,huang2017all,lee2021neural,wu2022learning,komaromi2024enhancing,sun2023htec,wu2023jump} and functional magnetic resonance imaging (fMRI) to observe the potential effects of physical exercise on brain activity in children with ADHD. Meanwhile, smart monitoring devices were used throughout the study to track physical activity levels and provide real-time feedback on the children's cognitive progress, helping researchers fine-tune interventions as needed.

Based on the theoretical framework of cognitive neuroscience and motor neuroscience~
\cite{wang2024cross,lee2024traffic,peng2025integrating,wang2024deep,wan2024image,jiang2020dualvd}, this study investigated the potential mechanisms through which physical exercise influences the executive function of children with ADHD. From the perspective of cognitive neuroscience, the improvement of executive function is closely related to the neural activity of the prefrontal cortex, which is one of the brain areas most significantly affected by physical exercise. Studies have shown that physical exercise can promote executive function by increasing blood flow and neuroplasticity in the prefrontal cortex. Studies have found that physical exercise can increase the synthesis and release of dopamine, thereby improving the attention and impulse control ability of children with ADHD. The theory of sports neuroscience further supports the positive effects of physical exercise on executive function. Sports activities such as coordination training can not only improve children's physical coordination ability but also enhance the brain's ability to handle complex tasks. This is because coordination training requires the brain to constantly adjust the movement of different parts of the body, thereby promoting the brain's task-switching ability and the development of working memory. In this study, smart monitoring devices (as shown in Figure \ref{fig7}) were used to track real-time brain activity and physical responses during exercise, providing valuable insights into how different exercise types affect neural activation. Therefore, the design of this study not only considers the overall cognitive effect of physical exercise but also compares different types of exercise interventions, striving to comprehensively explore its potential impact on children with ADHD.

Through this study, it is expected to provide new evidence support for non-drug interventions for children with ADHD. Compared with traditional drug treatments, physical exercise has broad application prospects as an intervention method with no side effects and is easy to promote. The integration of smart monitoring during the intervention allows for the collection of real-time feedback, enabling personalized and more effective intervention strategies. This study also points out the direction for future research. Although existing studies have preliminarily confirmed the positive effects of physical exercise on the cognitive function of children with ADHD, there are still many mysteries. For example, do different types of physical activities have different cognitive effects? Are the effects of long-term physical exercise persistent? Future studies can further explore the answers to these questions through larger-scale randomized controlled trials and combined with more neuroimaging techniques. Smart monitoring systems in future research will also help track the dynamic effects on cognitive and motor functions more accurately. In summary, physical exercise, as a potential non-drug intervention, can effectively improve the executive function and motor performance of children with ADHD. This study systematically evaluated physical exercise's cognitive and motor effects on children with ADHD by designing two different PA intervention programs and looks forward to providing new insights and empirical support in this field.

\section{Related Work}
In the study of children with ADHD, executive function deficits have always been the focus of research. In recent years, with the significant achievements of physical activity (PA) in improving cognitive function, more and more studies have begun to explore its potential impact on the executive function of children with ADHD.

\subsection{Effects of PA on cognitive function in healthy children}

A large number of studies ~\cite{nejati2023effect,rastikerdar2023developmental,zelazo2023reconciling} have shown that regular physical exercise has a significant promoting effect on the cognitive function of healthy children, especially in executive functions such as attention, working memory, inhibitory control and task switching. A study conducted by Ferrer et al. ~\cite{ferrer2022can} found that aerobic exercise can significantly improve the concentration ability and cognitive flexibility of school-age children. This study used cognitive task tests and found that children who participated in physical exercise performed significantly better than the control group in executive function tests after the intervention. Similarly, Davis et al. ~\cite{davis2007effects} conducted a 12-week aerobic exercise training on children and found that physical activity can not only improve cognitive function, but also promote academic performance.
In addition to aerobic exercise, coordination training and strength training have also been shown to have a positive effect on cognitive function. For example, the study of Diamond and Lee ~\cite{diamond2011interventions} pointed out that sports training involving complex action planning and coordination can help improve children's working memory and self-control ability. These studies provide an important theoretical basis for the research of this article, that is, physical exercise has a significant promoting effect on the cognitive function of healthy children.

\subsection{Effects of PA on cognitive and motor abilities in children with ADHD}
Compared with healthy children, children with ADHD have significant deficits in executive function and motor ability. To explore the intervention effect of PA on children with ADHD, many studies ~\cite{augusto2023exercise,keller2023hierarchical,shahaeian2023role} have begun to use physical exercise as one of the non-drug intervention methods. A study conducted by Smith et al. ~\cite{smith2013pilot} found that children with ADHD who participated in physical exercise showed significant improvements in task switching and working memory tests. The study adopted a randomized controlled trial design and divided children with ADHD into an exercise intervention group and a control group. The study by Pontifex et al. ~\cite{pontifex2013exercise} showed that children with ADHD showed better motor control and body coordination after participating in aerobic exercise. This study pointed out that children with ADHD usually show poor coordination in motor tasks, and physical exercise can improve these motor deficits by enhancing neuroplasticity in the brain.
Although existing studies have shown that PA has a certain positive effect on the cognitive function and motor performance of children with ADHD, existing studies still have some limitations. First, the sample size of most studies is small and the intervention time is short, making it difficult to fully evaluate the long-term effects of physical exercise on children with ADHD. Second, there are few comparative studies on different types of physical activities, and it is not clear which type of physical activity is most effective in improving cognitive function in children with ADHD. 

\subsection{Limitations of existing intervention studies}
Existing intervention studies still have some shortcomings in design and implementation ~\cite{bisset2023practitioner,ramey2023gaps,eltyeb2023attention,phalke2023auditory,nguyen2023depth}. First, many studies lack long-term follow-up, making it difficult to determine the long-term effects of physical exercise. For example, although many studies have shown that short-term physical activity can improve the executive function of children with ADHD, it is unclear whether its effects can last for months or even years after the end of the intervention. In addition, many studies use relatively simple intervention methods, while ignoring the potential effects of diverse forms of exercise on cognitive function. Therefore, future studies should try to combine multiple forms of exercise to explore their different effects.
Second, existing studies also have certain limitations in sample selection. Most intervention studies focus on a smaller sample of children with ADHD and lack systematic control of variables such as gender, age, and severity of the disease. Because children with ADHD have great differences in cognitive and motor abilities, future studies should further consider the impact of individual differences on intervention effects, especially designing more personalized intervention programs for children with different subtypes of ADHD.

\subsection{Potential mechanisms of physical exercise in improving executive function}
The potential mechanism of physical exercise in improving cognitive function is a complex multifactorial process involving multiple factors such as brain structure, neurotransmitter regulation, and cognitive processes. Studies ~\cite{ferlito2023acute,dastamooz2023efficacy,huang2023chronic,li2023effect,zheng2024triz,liu2025eitnet,an2023runtime,yuan2025gta,ren2025iot,zhang2024deep,liu2024dsem} have shown that physical exercise can enhance neuroplasticity by increasing blood flow in the brain and promoting neuron generation, especially in the prefrontal cortex area closely related to executive function. The prefrontal cortex is the core area of the brain responsible for high-level cognitive tasks, including planning, decision-making, task switching, impulse inhibition, etc. Physical exercise can improve executive function by enhancing neural activity in this area.
In addition, physical exercise can also regulate neurotransmitters in the brain, such as dopamine and norepinephrine. These neurotransmitters are closely related to the core symptoms of ADHD, especially the functional defects of the dopamine system, which are considered to be the main cause of executive function deficiency in children with ADHD. Through physical exercise, the dopamine level in the brain of children with ADHD may be increased, thereby improving their attention and impulse control ability.

Although studies have shown that physical exercise has the potential to improve the cognitive and motor abilities of children with ADHD, this study is unique in many aspects. First, this paper designed two different sports activity intervention programs, aerobic exercise and coordination training, in an attempt to compare the different effects of different types of sports activities on the cognitive function of children with ADHD. Secondly, this study used a variety of assessment tools, including cognitive task tests, motor performance assessments, and neuroimaging techniques (such as EEG and fMRI)~\cite{li2024ltpnet,zhao2025short,wang2024intelligent,zhou2024optimization,xi2024enhancing,wang2024usin}, to provide multi-dimensional data support for a comprehensive assessment of the effects of physical exercise on children with ADHD.
In summary, although existing studies have preliminarily shown that physical exercise has a positive effect on the abilities of children with ADHD, there are still many unresolved issues that need further exploration. The design of this study will make up for the shortcomings of existing research and aims to provide new theoretical basis and empirical support for non-drug interventions for children with ADHD.

\section{Method}

\begin{table*}[htbp]
\centering
\caption{Anthropometrical and fitness variables}
\resizebox{\linewidth}{!}{
\begin{tabular}{|l|c|c|c|c|c|}
\hline
\textbf{Variable} & \textbf{Experimental Group 1 (n = 13)} & \textbf{Experimental Group 2 (n = 14)} & \textbf{Control Group (n = 16)} & \textbf{F(1,43)} & \textbf{p} \\ \hline
Age (years)       & 9.2 (1.3)                             & 9.6 (1.6)                             & 9.5 (1.4)                       & 0.302            & 0.741      \\ \hline
Body mass index (BMI) (kg/cm²) & 18 (2.4)                    & 18.4 (2.4)                            & 18.7 (2.4)                      & 0.197            & 0.822      \\ \hline
Activities in a sports club (h/week) & 1.3 (1.7)            & 0.9 (1.5)                             & 1.1 (1.2)                       & 0.277            & 0.759      \\ \hline
Leisure sports (h/week) & 4.8 (5.2)                        & 4.2 (2.8)                             & 4.5 (2.5)                       & 0.092            & 0.913      \\ \hline
\end{tabular}}
\label{tab1}
\end{table*}

\subsection{Study design}
This study adopts an interventional experimental design to explore the effects of different types of physical activity (PA) on executive function (EF) in children with attention deficit hyperactivity disorder (ADHD). Participants include two experimental groups (EG1 and EG2) and a control group (CG), each of which receives 12 weeks of intervention training or maintains daily activities. By measuring cognitive function and motor performance multiple times, this study will comprehensively analyze the short-term and long-term effects of different types of physical activities on children with ADHD.
In the research design, the experiment adopts a random allocation method to assign children to different groups to ensure the balance of baseline characteristics between groups. The balance of each group in variables such as age and body mass index (BMI) was confirmed by statistical analysis, and the specific data are shown in Table \ref{tab1}.

All participants met the diagnostic criteria for ADHD in the Diagnostic and Statistical Manual of Mental Disorders (DSM-IV) and were confirmed by professional pediatricians. To ensure the independence of the intervention program, all participants did not receive any drug treatment or systematic behavioral intervention before the start of the study. Children with serious health problems, learning disabilities, or other neurodevelopmental disorders were excluded from the study to ensure the consistency of participants on a health and cognitive basis. In addition, any children with significant motor impairment or motor discomfort were also excluded to ensure that the study intervention was suitable for all participants. The age range of the participants was 7-10 years old, and the gender ratio was balanced. The baseline data are shown in Table \ref{tab1}. The balance of the groups in variables such as age and BMI was confirmed by analysis of variance (ANOVA), indicating that the groups were comparable at the beginning of the intervention. The study randomly assigned participants to the experimental group and the control group. There were 13 children in the EG1 group, 14 children in the EG2 group, and 16 children in the control group. The main purpose of randomization was to control potential bias factors and ensure the objectivity of the intervention effect. This study used standardized measurement tools to assess the working memory and motor performance of the participants. These tools are widely used in the fields of cognitive science and sports science and can provide reliable data support for evaluating the intervention effect.

\begin{table*}[htbp]
\centering
\caption{Exercise Program for EG1.}
\resizebox{\linewidth}{!}{
\begin{tabular}{|c|l|l|c|c|l|}
\hline
\textbf{Week} & \textbf{Training Content} & \textbf{Training Goals} & \textbf{Duration (min)} & \textbf{Difficulty Level} & \textbf{Main Equipment} \\ \hline
1 & \makecell[l]{Parent information introduction \\ and pre-test}  & \makecell[l]{Understand parent and child situation,\\ collect baseline data} & 20 & Low & None \\ \hline
2 & \makecell[l]{Catching, throwing,\\ and bouncing training} & Improve hand-eye coordination & 20 & Medium & Ball \\ \hline
3 & Balance training & \makecell[l]{Improve balance \\ and body control} & 20 & Medium & Balance board \\ \hline
4 & Acrobatics training &\makecell[l]{Enhance body flexibility \\ and coordination}  & 20 & High & Acrobatics tools \\ \hline
5 & Target throwing training & \makecell[l]{Improve accuracy \\ and concentration} & 20 & Medium & Target \\ \hline
6 & Tennis training & \makecell[l]{Improve reaction speed \\ and hand-eye coordination}  & 20 & Medium & Tennis racket \\ \hline
7 & Slacklining & \makecell[l]{Enhance body control \\ and focus} & 20 & High & Slackline \\ \hline
8 & Juggling & \makecell[l]{Improve coordination \\ and multitasking ability} & 20 & High & Juggling tools \\ \hline
9 & Beach volleyball, handball &\makecell[l]{Improve teamwork \\ and reaction speed}  & 20 & Medium & Volleyball \\ \hline
10 & Juggling training & \makecell[l]{Improve complex motor \\ coordination ability}  & 20 & High & Juggling tools \\ \hline
11 & Slacklining & Enhance balance and focus & 20 & High & Slackline \\ \hline
12 & Coordination training & Improve overall body coordination & 20 & Medium & Multiple equipment \\ \hline
13 & Ball catching training & Improve hand-eye coordination & 20 & Medium & Ball \\ \hline
14 & Post-test 2 & Test the effectiveness of the intervention & 20 & Low & None \\ \hline
\end{tabular}}
\label{tab2}
\end{table*}

\begin{table*}[htbp]
\centering
\caption{Exercise Program for EG2.}
\resizebox{\linewidth}{!}{
\begin{tabular}{|c|l|l|c|c|l|}
\hline
\textbf{Week} & \textbf{Training Content} & \textbf{Training Goals} & \textbf{Duration (min)} & \textbf{Difficulty Level} & \textbf{Main Equipment} \\ \hline
1 & \makecell[l]{Parent information introduction\\ and pre-test} & \makecell[l]{Understand parent and \\ child situation, collect baseline data} & 20 & Low & None \\ \hline
2 & \makecell[l]{Sports competition\\ - relay race} & \makecell[l]{Improve body coordination \\ and participation} & 20 & Medium & Relay baton \\ \hline
3 & Swimming training & \makecell[l]{Improve body coordination \\ and endurance} & 20 & Medium & Swimming pool \\ \hline
4 & Swimming training & \makecell[l]{Improve body coordination \\ and endurance} & 20 & Medium & Swimming pool \\ \hline
5 & Wrestling games & \makecell[l]{Increase strength \\ and reaction ability} & 20 & High & Mats \\ \hline
6 & Climbing training & \makecell[l]{Enhance body flexibility \\and upper body strength} & 20 & High & Climbing wall \\ \hline
7 & Climbing training & \makecell[l]{Enhance body flexibility \\ and upper body strength} & 20 & High & Climbing wall \\ \hline
8 & Orienteering training & \makecell[l]{Improve endurance and \\ teamwork skills} & 20 & High & Orienteering equipment \\ \hline
9 & \makecell[l]{Sports competition\\ - relay race} & \makecell[l]{Improve body coordination \\ and participation} & 20 & Medium & Relay baton \\ \hline
10 & Gymnastics training & \makecell[l]{Improve body coordination \\ and flexibility} & 20 & Medium & Gymnastics equipment \\ \hline
11 & \makecell[l]{Gymnastics - \\ trampoline training} & \makecell[l]{Improve reaction speed\\ and balance} & 20 & Medium & Trampoline \\ \hline
12 & Track and field training &\makecell[l]{Strengthen endurance \\ and explosive power}  & 20 & High & Track field \\ \hline
13 & Track and field training &\makecell[l]{Improve speed \\and coordination}  & 20 & High & Hurdles \\ \hline
14 & Post-test 2 & \makecell[l]{Test the effectiveness \\of the intervention} & 20 & Low & None \\ \hline
\end{tabular}}
\label{tab3}
\end{table*}

\subsection{Measuring tools}
This study used standardized measurement tools to assess participants' working memory and motor performance. These tools are widely used in the fields of cognitive science and sports science and can provide reliable data support for evaluating intervention effects.

\textbf{1) Working memory measurement.}
\begin{itemize}
    \item Verbal Working Memory (Verbal WM): The digit span test (both forward and reverse) and the letter-number sequencing task in HAWIK-IV were used. The digit span test requires participants to repeat a set of numbers to assess their short-term memory ability. The letter-number sequencing task further tests the complexity of working memory, requiring participants to rearrange a mixed sequence in the order of letters and numbers to assess their memory retention and processing ability.
    \item Visuospatial Working Memory (Visuospatial WM): The Corsi square tapping test was used. This task requires children to tap squares in a specified order and then repeat in the same order or reverse order to assess their spatial memory retention and processing ability.
\end{itemize}

\textbf{2) Motor ability measurement.}
\begin{itemize}
    \item KTK test (Körperkoordinationstest für Kinder): The KTK test is used to assess children's motor coordination ability. The test covers manual dexterity, catching and aiming, static and dynamic balance, and can comprehensively evaluate children's sports performance.
\end{itemize}

\subsection{Intervention options}
The intervention period of this study was 12 weeks, with physical activity training conducted twice a week, each lasting 20 minutes. Experimental Group 1 (EG1) and Experimental Group 2 (EG2) received different types of physical activity interventions, while the control group (CG) did not perform any designated physical activity training during this period.

\textbf{1) Intervention Program of EG1.}
The intervention program of EG1 mainly focused on improving catching, aiming, balance and manual dexterity. Specific training content included catching, throwing, balance training and acrobatics. Table \ref{tab2} describes the weekly training schedule, its goals and duration in detail. The training projects involved hand-eye coordination, balance ability, etc., aiming to improve children's executive function and cognitive performance through these complex motor coordination tasks.

\textbf{2) Intervention Program of EG2.}
The intervention program of EG2 focused on participating in less physically demanding but diverse sports activities, such as swimming, handball, track and field training, etc. Table \ref{tab3} lists the weekly training content, goals and equipment requirements in detail. This group of activities aims to enhance children's basic motor skills and improve collaboration and interactivity in group activities.
To analyze the short-term and long-term effects of different physical activity interventions on the executive function and motor performance of children with ADHD, this study used one-way analysis of variance (ANOVA) to compare the changes between the experimental group and the control group before and after the intervention. By comparing the baseline data of different groups, it was ensured that all participants had similar characteristics before the intervention. The significance of the intervention effect was based on a p value less than 0.05. If the p value reached this standard, it indicated that the intervention had a significant effect on the executive function and motor performance of children.

The significance level was set at p < 0.05. If the ANOVA results showed that the p value was less than 0.05, it indicated that physical activity had a significant improvement in the executive function of children with ADHD. This study not only evaluated the short-term effects by measuring working memory and motor performance multiple times, but also examined the long-term effects after the 12-week intervention.

\section{Results Analysis}
This study aimed to evaluate the short-term and long-term effects of different types of physical activities. According to the previously designed experimental methods, we analyzed the short-term and long-term intervention effects of EG1, EG2, and control group (CG).

\subsection{Short-term effects}
After a 12-week short-term intervention, EG1 (specific skill training group) showed significant immediate improvements in the catching and aiming tasks. In the first measurement after the training, participants in EG1 showed significant improvements in catching and aiming accuracy and speed compared with participants in EG2 (low-demand exercise group) and CG (control group). The performance of the EG2 and CG groups in this task did not change significantly, indicating that specific hand-eye coordination skill training has a particularly significant effect on improving this task.

\begin{figure}
    \centering
    \includegraphics[width=1\linewidth]{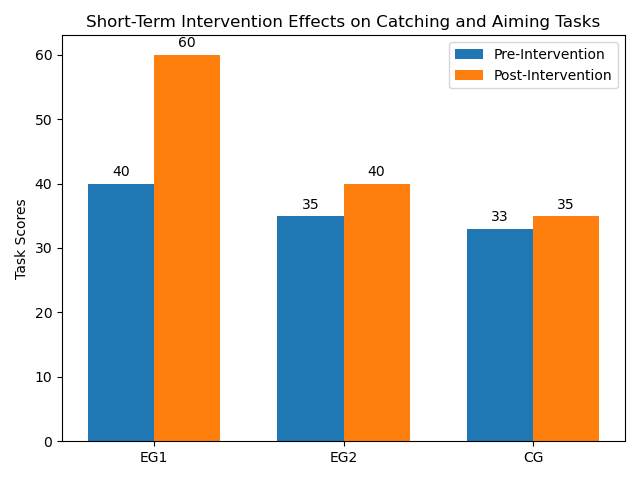}
    \caption{Short-term intervention effects of capture and targeting tasks.}
    \label{fig1}
\end{figure}

As shown in Figure \ref{fig1}, EG1 scored significantly higher in the catching and aiming tasks after the intervention than before the intervention, while the difference between EG2 and CG was not significant.

In the balance and manual dexterity tasks, the short-term intervention failed to bring significant between-group differences. Although participants in both the specific skill training group (EG1) and the low-demand exercise group (EG2) showed some improvement after the 12-week training period, these changes did not reach statistical significance when compared to the control group (CG). One possible explanation for this lack of significant improvement could be the relatively short duration of the intervention. Given that balance and manual dexterity require the integration of multiple motor skills and neural processes, it may take more time for these complex abilities to show noticeable improvement. Short-term interventions might not provide enough time for these processes to fully engage and result in measurable changes.

Additionally, the nature of the tasks themselves may contribute to the lack of significant improvement. Balance and manual dexterity involve not only physical coordination but also the ability to adjust motor output to changing environmental conditions, which requires more extensive and sustained practice over time. The intervention in this study might not have been intensive enough or specifically targeted these aspects of motor performance.

The results of the short-term intervention suggest that while specific skill training—such as tasks involving catching and aiming—can lead to relatively quick improvements in children's motor performance, it has limited effects on more complex skills like balance and manual dexterity. These findings highlight the need for a longer, more tailored intervention for tasks that involve intricate motor coordination and complex neural processes. Future studies could explore longer-term interventions or different types of exercise targeting these areas to better understand how to enhance these complex motor skills in children with ADHD.

\subsection{Long-term effects}
\begin{figure}
    \centering
    \includegraphics[width=1\linewidth]{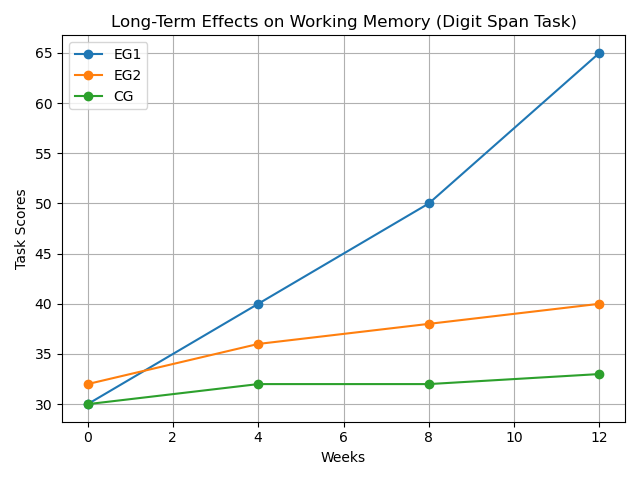}
    \caption{Long-term intervention effects of a digital photometry task.}
    \label{fig2}
\end{figure}

In terms of long-term intervention effects, the study found that all participants improved their performance on working memory tasks (WM), especially participants in EG1 showed more significant improvements. Specifically, EG1's working memory scores in digit span, vocabulary fluency, and letter-number sorting tasks showed significant group × time interaction effects. This shows that after 12 weeks of specific skill sports activity training, participants in EG1 not only improved their sports performance but also made significant progress in cognitive function, especially working memory.
The score of the digit span task improved significantly in the EG1 group, while the improvement in the EG2 and CG groups was smaller and did not reach the significance level. Figure \ref{fig2} shows the average scores and change trends of the three groups of participants before and after the intervention. It can be seen that EG1's working memory ability showed a significant improvement after the intervention.

Similarly, in the letter-number sorting task and vocabulary fluency task, participants in EG1 showed significant improvement, while the improvement in the EG2 and CG groups was relatively small. As shown in Figure 3, the long-term intervention effect of EG1 was significant on these two tasks, indicating that targeted physical training not only helps improve physical coordination, but also has a profound impact on cognitive function.
Although this study did not directly measure inhibitory control tasks, according to existing literature and theoretical hypotheses, physical activities, especially training involving complex coordination and hand-eye coordination tasks, may have potential benefits for inhibitory control in children with ADHD. Future studies can further design and evaluate the specific effects of physical intervention on inhibitory control.

\begin{figure}
    \centering
    \includegraphics[width=1\linewidth]{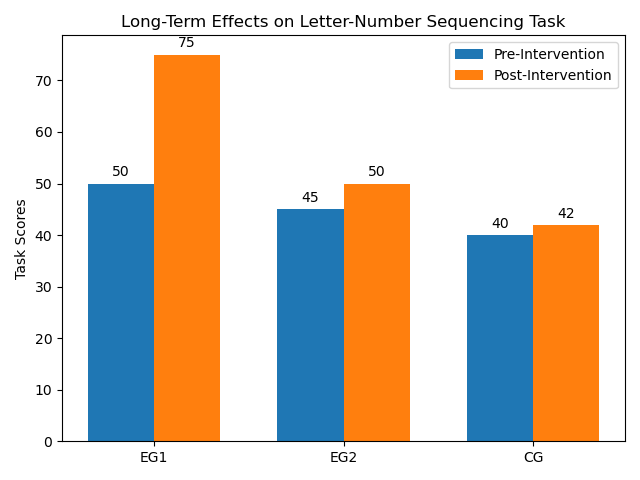}
    \caption{Long-term intervention effects on a letter-number sequencing task.}
    \label{fig3}
\end{figure}

According to the results of this study, long-term physical activity intervention has a significant positive effect on cognitive function, especially working memory, in children with ADHD. In particular, participants in the EG1 group not only improved their motor skills through specific skill training (such as catching, aiming, and manual dexterity training), but also showed significant improvements in cognitive functions such as working memory. This finding suggests that more targeted physical activity intervention can bring greater cognitive benefits than general physical activity.
The effect of short-term intervention is more limited, especially in the balance and manual dexterity tasks. This may be related to the short intervention time or the complexity of these tasks themselves. Compared with simple, direct motor tasks such as catching and aiming, balance and dexterity involve more complex physical coordination abilities, so longer training may be required to achieve significant progress.

This study shows that long-term specific sports activity intervention can significantly improve the working memory and motor performance of children with ADHD, especially in specific skills such as catching, aiming, and manual dexterity. However, short-term intervention has limited effect on improving balance ability. The study also suggests that future intervention studies should further examine the potential impact of physical activities on children's inhibitory control ability, as well as the effects of different intervention times on different executive functions and motor skills.

\subsection{Subtypes of ADHD and the Need for Further Differentiation in Intervention Effects}
This study explores the impact of physical activity (PA) interventions on the executive function and motor performance of children with ADHD. A 12-week intervention was designed, involving different types of PA, including a specific skill training group (EG1), a low-demand exercise group (EG2), and a control group (CG). Participants were randomly assigned to one of the three groups, and the effects of the interventions were assessed using various cognitive and motor tasks, such as catching and aiming tasks, balance and manual dexterity tests, and working memory tasks. The study tracked participants’ progress through performance evaluations before and after the intervention.

Additionally, the study utilized smart monitoring devices to track physical activity levels, allowing for real-time feedback on performance and activity. These devices helped ensure that the interventions were consistently followed, and they provided a deeper understanding of how different types of physical activities affected the children’s cognitive functions.

However, it is important to note that the study does not account for different subtypes of ADHD, such as the inattentive or hyperactive-impulsive subtypes. This distinction could be valuable, as ADHD is a heterogeneous disorder, and the interventions might have varying effects depending on the subtype of ADHD a child exhibits. Differentiating between these subtypes could potentially reveal more targeted and nuanced effects of the interventions across various ADHD profiles. Future studies would benefit from incorporating this differentiation to assess the effectiveness of PA interventions for each ADHD subtype, offering insights into more personalized and effective treatment strategies.

\section{Discussion}

The results of this study provide strong empirical support for our understanding of the effects of physical activity on executive function in children with ADHD. Through the analysis of short-term and long-term intervention effects, we found that specific skill training has a significant positive effect on working memory and motor performance in children with ADHD, especially in catching and aiming tasks, working memory, etc. Compared with low-demand physical activities, targeted physical training can provide children with more significant cognitive and physical benefits.

\subsection{Short-term effects}
During the short-term intervention phase, EG1 showed significant improvements, especially in the catching and aiming tasks. In contrast, EG2 and the CG did not show significant short-term changes. This result suggests that specific sports skill training can quickly improve the motor skills of children with ADHD, especially hand-eye coordination. This may be because the training program of EG1 focuses on repeated precision movements, such as catching and throwing, which require immediate reactions and concentration and can effectively improve performance in the short term.

\begin{figure}
    \centering
    \includegraphics[width=1\linewidth]{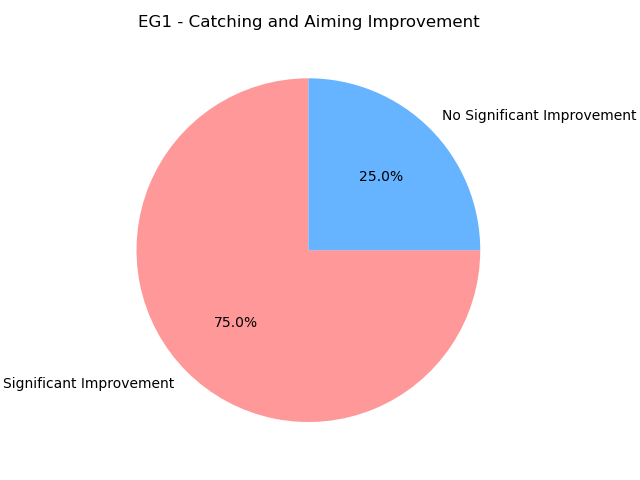}
    \caption{EG1-Catching and Aiming Improvement.}
    \label{fig4}
\end{figure}

\begin{figure}
    \centering
    \includegraphics[width=1\linewidth]{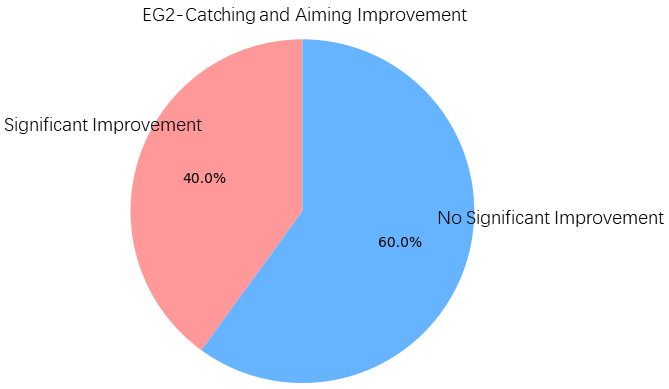}
    \caption{EG2-Catching and Aiming Improvement.}
    \label{fig5}
\end{figure}

\begin{figure}
    \centering
    \includegraphics[width=1\linewidth]{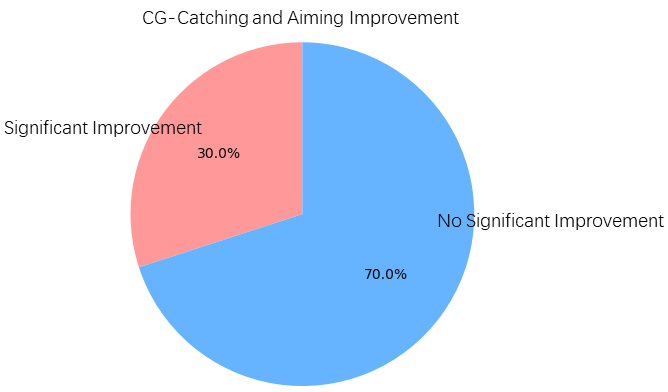}
    \caption{CG-Catching and Aiming Improvement.}
    \label{fig6}
\end{figure}

In Figures \ref{fig4}, \ref{fig5}, and \ref{fig6}, we use a pie chart to show the proportion of children in the three groups who showed significant improvement in the catching and aiming tasks. The chart can more intuitively see that a larger proportion of children in the EG1 group showed significant short-term progress, while the EG2 and CG groups were relatively low.

As can be seen from Figures, 75\% of the children in the EG1 group showed significant improvement in the catching and aiming tasks, while only 40\% and 30\% of the children in the EG2 and CG groups showed similar improvements, respectively. This further verifies that the effect of specific skill training is more significant.

\subsection{Long-term effects}
In terms of long-term intervention, the study showed that all participants improved their working memory (WM) performance, but the improvement in EG1 was particularly prominent. The EG1 group showed a significant group × time interaction effect in the digit span task, vocabulary fluency task, and letter-number sorting task, indicating that long-term physical activity can not only improve the motor ability of children with ADHD, but also significantly improve their cognitive ability, especially working memory performance.

These results are consistent with existing cognitive neuroscience research~\cite{dong2024design,zhuang2020music,wang2024recording,Wang2024Theoretical,lyu2024optimized,guo2024construction}, that is, physical activity, especially complex motor tasks, can stimulate the activity of the prefrontal cortex, thereby improving executive function. The training content of the EG1 group focused on tasks such as coordination and hand-eye coordination, which may have stimulated these brain areas more directly. In contrast, the improvements in the EG2 and CG groups were relatively small, which also verified the unique advantages of targeted intervention programs for cognitive improvement.

\subsection{Potential impact of inhibitory control}
Although this study did not directly measure inhibitory control ability, based on existing literature, it is speculated that physical activities, especially complex hand-eye coordination training, may have a positive effect on inhibitory control. Previous studies have shown that various aspects of executive function, such as working memory, task switching, and inhibitory control, are interrelated and can influence each other. For instance, activities that improve working memory and task switching might also support improvements in impulse control by reinforcing the brain's ability to regulate attention and control actions. Future studies should further design experiments to explore the direct effects of physical activities on inhibitory control in children with ADHD and combine neuroimaging techniques, such as functional MRI or EEG, to verify changes in specific brain regions associated with these cognitive processes.

There are still some limitations to this study. First, although the intervention period was 12 weeks, for more complex skills such as balance and dexterity, longer training may be required to see significant changes. The 12-week duration may not have been enough to induce measurable improvements in these skills, which involve coordination between various neural systems. Additionally, while the study focused on working memory and motor skills, the effects of other executive functions, such as inhibitory control, were not directly assessed. Future studies can combine more dimensional cognitive function measurements to improve knowledge in this field, offering a more comprehensive understanding of how physical activity influences executive function in children with ADHD.

Another limitation is the relatively small sample size. Although random grouping reduces potential bias, larger studies may increase the generalizability of the results and provide more robust evidence for the effectiveness of physical activity interventions. Furthermore, all participants in this study were aged 7-10, and it would be valuable to expand the scope of future research to explore the different responses of children of various ages to physical activity interventions. This would help identify age-specific patterns and tailor interventions to the developmental stages of children with ADHD.

\section{Conclusion}
This study verified the effects of different types of physical activities on executive function and motor performance in children with ADHD through a 12-week intervention experiment. The results showed that specific skill training (such as catching and aiming) can quickly improve the hand-eye coordination ability of children with ADHD in the short term, while other motor tasks (such as balance and manual dexterity) did not show significant short-term changes. However, the effect of long-term intervention was more significant, especially in the working memory task. The EG1 group showed a significant group × time interaction effect, indicating the improvement of specific skill training on cognitive function.
Although the improvements in the EG2 and CG groups were relatively small, this does not negate the overall positive effect of physical activity on children with ADHD. On the contrary, this suggests that targeted physical activity intervention may be more effective and can enhance children's cognitive and motor abilities through complex coordination tasks and precise movements. Future studies should continue to explore the effects of different types of physical activities on different age groups and different cognitive dimensions, especially by combining neuroimaging technology to further understand how physical activities promote the improvement of executive function in children with ADHD.

In summary, this study provides strong evidence for non-drug intervention for children with ADHD, indicating that long-term specific skill sports activities can significantly improve the working memory and motor skills of children with ADHD. This provides schools and families with feasible intervention methods, supporting physical activity as an effective way to assist in the treatment of children with ADHD.

\section*{Author Contributions}
\textbf{Liwen Lin:} Conceptualization, Methodology, Writing–original draft;
\textbf{Nan Li:} Conceptualization, Writing-review \& editing;
\textbf{Shuchen Zhao:} Methodology, Writing-review \& editing.

% \section*{Acknowledgement}

\section*{Data availability}
The data that support the findings of this study are available on request from the corresponding author. The data are not publicly available due to privacy or ethical restrictions.

\section*{conflicts of interest}
The authors declare that they have no known competing financial interests or personal relationships that could have appeared to influence the work reported in this paper.

\bibliographystyle{model1-num-names}
\bibliography{cas-refs}


\end{document}